\documentclass[twocolumn,showpacs,preprintnumbers,amsmath,amssymb,prb]{revtex4}

\usepackage{graphicx,psfrag}
\usepackage{dcolumn}
\usepackage{bm}

\begin{document}

\preprint{APS/123-QED}

\title{On superconductivity in armchair carbon nanotubes.}

\author{A. S\'{e}d\'{e}ki}
\author{L. G. Caron}	
\author{C. Bourbonnais}
\affiliation{
Centre de recherche sur les proprit\'{e}s
\'{e}lectroniques de mat\'{e}riaux avanc\'{e}s., D\'{e}partement de Physique,\\
Universit\'{e} de Sherbrooke, Sherbrooke, Qu\'{e}bec, Canada J1K 2R1.
}
 
\date{\today}

\begin{abstract}
We use the momentum space renormalization group to study the influence of
phonons and the Coulomb interaction on the superconducting response function
of armchair single-walled nanotubes. We do not find superconductivity in
undoped single nanotubes. When doped, superconducting fluctuations can
develop because of the phonons but remain small and are easily destroyed by
the Coulomb interaction. The origin of superconductivity in ropes of
nanotobes is most likely an intertube effect. Projections to zig-zag
nanotubes indicate a more favorable disposition to superconducting
fluctuations.
\end{abstract}

\maketitle

Since their discovery by Ijima\cite{ijima} in 1991, carbon nanotubes have
attracted a lot of interest due to their unusual geometry and their
structural and electronic properties\cite{Dressel}. Band calculations \cite
{theorie} (confirmed by experiments\cite{Dunlap,exp}) have stressed the
one-dimensional (1D) character of single-wall carbon nanotubes (SWCNT).

Recently Kociak {\it et} {\it al.}\cite{Kociak}{\it \ }reported measurements
on ropes of SWCNT making low-resistance contacts to non-superconducting
(normal) metallic pads, at low voltage and at temperatures down to $70{\rm mK}$. The preparation technique they have used yields armchair nanotubes\cite
{preparation}. Their results show signs of superconductivity below $0.55{\rm K}$. The authors predict a purely electronic mechanism. The question we
address here is wether or not 1D superconducting fluctuations can exist in
single SWNT and if phonons can play any role.

\begin{figure}
\psfrag{E}{$E$}
\psfrag{Fermi level}{Fermi level}
\psfrag{(+)}{($+$)}
\psfrag{(-)}{($-$)}
\psfrag{k}{$k$}
\includegraphics[scale=1.8]{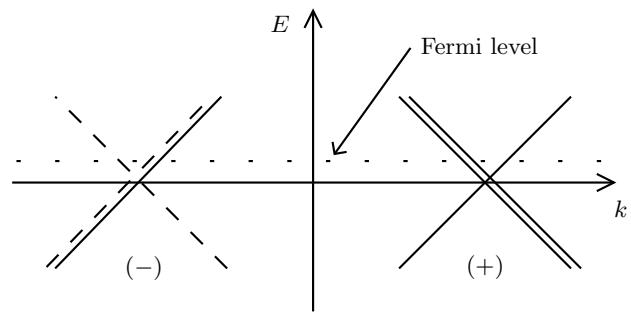}
\caption{\label{bande} Energy bands of an armchair nanotube near the Fermi energy. The
double and single lines refer to each of the overlapping bands. The Fermi
level is at the crossing point in the undoped tube and shifts when doped.}
\end{figure}

To achieve this, we perform perturbative renormalization group (RG)
calculations\cite{Solyom,Bourbonnais} to analyze the low energy
behavior of a ($n,n$) armchair nanotube. The electronic and phonon parts 
$H_{0}$ and the electron-phonon contribution $H_{e-ph}$ of the Hamiltonian
were described in a previous publication\cite{sedeki}. We added the Coulomb
interaction $H_{e-e}$\cite{Egger}. The electron wavefunctions come from a
nearest-neighbor tight-binding model using a{ \ }$\pi _{z}$ orbital on
each carbon atom of a sheet of graphene which is rolled up in the proper way
to generate an armchair nanotube\cite{Jishi}. In order to build an effective
Hamiltonian ($H=H_{0}+H_{e-e}+H_{e-ph}$) for low temperatures, we discard
all bands that do not intersect the Fermi level. We then linearize the band
energies around the Fermi level, $\varepsilon _{\gamma ,p}(k')=(-1)^{\gamma +1}pv_Fk'$ where $k^{\prime }=(k-pk_F)$, $\gamma
=1,2$ is the band index, $v_F$ is the Fermi velocity, $k_F$ is the Fermi
momentum, and $p=\pm $ is the sign of $k$. This is shown in Fig.~\ref{bande}. The Fermi
level lies exactly at the crossing point for a half-filled band. This is the
situation we shall first examine. 
In the g-ology approach to the RG, both
the direct\ electron-electron and the phonon mediated interactions give rise
to twelve independent scattering amplitudes $g_{i,e-e}^{(j)}$ and $
g_{i,ph}^{(j)}$ ($i=1,2,4$ and $j=1,..,4$). Here we have adopted the
notation of Krotov \cite{Krotov}. The index $i$ refers to the momentum
branches ($p$) and $j$ to the velocity branches (sign of the electron
velocity = $(-1)^{\gamma +1}p$) such that 1 = interbranch backscattering, 2
= interbranch forward scattering, 3 = umklapp scattering, and 4 =
intrabranch forward scattering. There are no momentum branch Umklapp
processes since $4k_{F}=8\pi /3a$ in not equal to a reciprocal lattice
vector, where $a$ is the tube's unit cell length. The coupling constants $%
g_{i,e-e}^{(j)}$ were calculated by summing over all tube sites using a
Coulomb interaction of the form used by Egger {\it et al.}\cite{Egger}: 
\begin{eqnarray}
g_{1,e-e}^{(1)}&=&g_{1,e-e}^{(3)}=b\text{, } g_{ 1,e-e}^{(2)}=g_{1,e-e}^{(4)}=b',\nonumber \\
g_{2,e-e}^{(1)}&=&g_{2,e-e}^{(3)}=g_{4,e-e}^{(1)}=g_{4,e-e}^{(3)}=u'\text{,}\label{ge-e} \\
g_{ 2,e-e}^{(2)}&=&g_{2,e-e}^{(4)}=g_{4,e-e}^{(2)}=g_{4,e-e}^{(4)}=u\text{,}\nonumber
\end{eqnarray}
where $u$ $(u^{\prime })$ and $b$ $(b^{\prime })$ are related to the
strength of the bare Coulomb interaction. The ratios $(u/b,b^{\prime
}/b,u^{\prime }/u)$ vary between $(2,0.005,0.01)$ and $(20,0.01,0.002)$ when
going from a strongly\ screened to an unscreened interaction.
The phonon mediated electron-electron coupling constants in the
non-adiabatic regime\cite{Caron}, $g_{i,ph}^{(j)}$, are given by\cite{sedeki}: 
\begin{equation}
g_{ i,ph}^{(j)}\equiv -{2 \over \omega _q}g_{e-ph,\gamma}(q)g_{e-ph,\gamma'}(-q)\label{ge-ph}
\end{equation}
where{ \ }$g_{e-ph,\gamma}(q)$ is the electron-phonon interaction with each
of the two electrons involved and $\omega _{q}$ is the phonon frequency. One
has $q \approx 2k_{F}$ for backward scattering and $q \approx 0$ 
for forward scattering. These parameters are valid for
temperatures or energies $(k_{B}=1)$ smaller than the phonon energy.

The symmetry of the electronic functions{ \ }\cite{Jishi} leads to the
following sign constraints for the $g_{i,ph}^{(j)}$ 
\begin{eqnarray}
g_{1,ph}^{(1)}=-g_{1,ph}^{(3)}=g_{1}<0\text{, }g_{1,ph}^{(2)}=0\text{, }g_{1,ph}^{(4)}<0,\nonumber \\ 
g_{2,ph}^{(1)}=-g_{2,ph}^{(3)}=g_{4,ph}^{(1)}=-g_{4,ph}^{(3)}=g_2<0,\nonumber \\ 
g_{2,ph}^{(2)}=-g_{2,ph}^{(4)}=-g_{4,ph}^{(2)}=g_{4,ph}^{(4)}=g_4<0.\label{phonon}
\end{eqnarray}
We then perform a one loop RG calculation, which is a generalization of the
one band calculation \cite{Solyom}\cite{Bourbonnais}, by considering only
the diagrams presenting a logarithmic divergence with temperature, that is
with electrons on different velocity branches. The result of this procedure
yields the same flow equations for the coupling constants that were found by
Krotov {\it et} {\it al.}\cite{Krotov}. At half-filling, we use a two
cutoff approach\cite{Caron} with $W>\hbar \omega _{ph}$ where $W$ is the
bandwidth and the renormalization starting energy scale while $\hbar \omega
_{ph}$ is the phonon energy. For energy (or temperature) scales $%
E_{0}=We^{-\ell }>\hbar \omega _{ph}$, the phonons are adiabatic and the $
g_{e-ph,\gamma}$ renormalize only in the random phase approximation. When $E_{0}$
reaches $\hbar \omega _{ph}$, the phonons become non-adiabatic, the $%
g_{i,ph}^{(j)}$ become active and are added to the $g_{i,e-e}^{(j)}$.
Contrary to\ the usual case of a single band crossing the Fermi level, the
coupling constants in our case scale towards the strong coupling sector. An
analytical solution of the renormalization equations seems to be impossible
in the general case, which forces us to resort to a numerical solution.

The important physical properties of the system can be probed through its
various response functions. These have the coupling constants as input and
measure the relative importance of the underlying fluctuations. It is
important to realize that, because of the interband interaction terms ($%
g_{1}^{(2)},g_{2}^{(1)},g_{i}^{(3)},g_{4}^{(j)})$, the particle-hole pairs
in the Peierls channel and the particle-particle pairs in the Cooper channel
will be evolving in both bands. It is thus necessary to define the following
response functions in Matsubara-Fourier space 
\begin{eqnarray}
\chi_{\mu}^{\kappa,M}(\tilde{q})&=&-\int_0^{\beta}\int{d\tau dx \ }e^{-iqx+i\omega_m\tau }\nonumber \\
&&\times\langle O_{\mu}^{\kappa,M}(x,\tau )^{\dagger}\ O_{\mu }^{\kappa,M}(0,0)\rangle\text{,}
\end{eqnarray}
with $\kappa =\pm $, $M=\pm $, and $\tilde{q}=(q,\omega _{m})$. The Fourier
transforms $O_{\mu }^{\kappa ,M}(\tilde{q})$ are defined by 
\begin{equation}
O_{\mu }^{\kappa,M}(\tilde{q})={1\over \sqrt{2}}(O_{\mu }^{\kappa}(\tilde{q})+MO_{\mu }^{\kappa }(\tilde{q})^{\dagger}).
\end{equation}
In the Peierls channel, one defines 
\begin{eqnarray}
O_{\mu}^{\kappa}(q\approx2k_F)&=&{1\over\sqrt{L}}\sum_{k,\alpha,\beta}[\psi_{1,-,\alpha }^{\dagger}(k-q)\sigma_{\mu }^{\alpha \beta}\psi_{1,+,\beta}(k)\nonumber \\
&&+\kappa \psi_{2,-,\alpha }^{\ast}(k-q)\sigma _{\mu }^{\alpha \beta }\psi _{2,+,\beta}(k)]/2,
\end{eqnarray}
\begin{eqnarray}
O_{\mu}^{\kappa}(q \approx 0)&=&{1 \over \sqrt{L}}\sum_{k,\alpha,\beta}[\psi_{1,-,\alpha}^{\dagger}(k-q)\sigma _{\mu }^{\alpha \beta}\psi_{2,-,\beta}(k)\nonumber \\
&&+\kappa \psi_{2,+,\alpha}^{\dagger}(k)\sigma _{\mu }^{\alpha \beta}\psi_{1,+,\beta}(k-q)]/2,
\end{eqnarray}
in which $\mu =0$ stands for charge density (CDW) operators and $\mu
=1,2,3 $, for spin density (SDW) ones. Here $\sigma _{0}$ and $\sigma
_{1,2,3}$ are the identity and the $x$, $y$, $z$ Pauli matrices,
respectively, and $\psi _{\gamma ,p,\alpha }(k)$ annihilates an electron of
spin $\alpha $ in band $\gamma $ having momentum $k$ in branch $p$. We have introduced the parameter $M$ by which we differenciate between on-site and bond-order for charge and spin correlation functions as defined in Ref.~\onlinecite{Caron}. In the Cooper channel, one has 
\begin{eqnarray}
O_{\mu}^{\kappa}(q\approx0)&=&{1\over\sqrt{L}}\sum_{k,\alpha,\beta}\alpha\lbrack\psi_{1,-,\alpha }(-k+q)\sigma_{\mu }^{-\alpha,\beta}\psi_{1,+,\beta }(k)\nonumber \\
&&+\kappa \psi_{2,-,\alpha}(-k+q)\sigma _{\mu }^{-\alpha ,\beta }\psi_{2,+,\beta }(k)]/2,
\end{eqnarray}
\begin{eqnarray}
O_{\mu}^{\kappa}(q\approx2k_F)&=&{1\over\sqrt{L}}\sum_{k,\alpha,\beta}\alpha\lbrack\psi_{1,-,\alpha}(-k+q)\sigma_{\mu }^{-\alpha,\beta}\psi_{2,-,\beta}(k)\nonumber \\
&&+\kappa\psi_{1,+,\alpha}(-k+q)\sigma _{\mu }^{-\alpha,\beta}\psi_{2,+,\beta}(k)]/2,
\end{eqnarray}
where $\mu =0$ are singlet superconducting (SS) operators and $\mu =1,2,3$
are triplet superconducting (ST) ones. It is through these band-entangled
operators that our response functions are different from the ones of Krotov 
{\it et} {\it al.}\cite{Krotov} who only used the untangled $\left( O_{\mu
}^{+}\pm O_{\mu }^{-}\right) $ operators. Our definition of the response
functions leads to a fundamentally different behavior with temperature.

In order to calculate the evolution of the response functions with the
energy scale $E_{0}$ (or temperature), we introduce the auxiliary response
functions $\bar{\chi}_{\mu }^{\kappa ,M}$ defined through
\begin{equation}
{ \chi }_{{ \mu }}^{\kappa { ,M}}{ (\ell ,\tilde{q})}%
={ -}\frac{{ 1}}{{ \pi v}_{{ F}}}\int_{{ 0}}^{%
{ \ell }}{ \bar{\chi}}_{{ \mu }}^{\kappa { ,M}}%
{ (\ell }^{\prime }{ ,\tilde{q})d\ell }^{\prime }\text{, }
\end{equation}
where $\ell =\ln (W/E_{0})$. We deduce the following renormalization
equations 
\begin{eqnarray}
{d \over d\ell}\ln \bar{\chi}_{CDW}^{\kappa,M} &=&\bar{g}_2^{(2)}(\ell)-2\bar{g}_1^{(1)}(\ell)+\kappa(\bar{g}_2^{(3)}(\ell)-2\bar{g}_1^{(3)}(\ell )),  \nonumber \\
{d \over d\ell}\ln \bar{\chi}_{SDW}^{\kappa,M} &=& \bar{g}_{2}^{(2)}(\ell)+\kappa \bar{g}_{2}^{(3)}(\ell ),  \nonumber \\
{d \over d\ell }
\ln \bar{\chi}_{CDW'}^{\kappa,M} &=&{\bar{g}_{4}^{(2)}(\ell )-2\bar{g}_{4}^{(1)}(\ell )+\kappa (\bar{g}%
_{1}^{(2)}(\ell )-2\bar{g}_{2}^{(1)}(\ell ))}  \nonumber \\
&&{ +M[-\bar{g}_{4}^{(3)}(\ell )+\kappa (\bar{g}_{1}^{(3)}(\ell )-2%
\bar{g}_{2}^{(3)}(\ell ))],}  \nonumber \\
{d\over d\ell}{ \ln \bar{\chi}}_{{ SDW}^{\prime }}^{\kappa { ,M}} &=&%
{ \bar{g}_{4}^{(2)}(\ell )+\kappa \bar{g}_{1}^{(2)}(\ell )+M[\bar{g}%
_{4}^{(3)}(\ell )+\kappa \bar{g}_{1}^{(3)}(\ell )],}  \nonumber \\
{d \over d\ell}%
{ \ln \bar{\chi}}_{{ SS}}^{\kappa} &=&{ -\bar{g%
}_{2}^{(2)}(\ell )-\bar{g}_{1}^{(1)}(\ell )+\kappa (\bar{g}_{2}^{(1)}(\ell )+%
\bar{g}_{1}^{(2)}(\ell )),}  \nonumber \\
{d \over d\ell}%
{ \ln \bar{\chi}}_{{ TS}}^{\kappa} &=&{ -\bar{g%
}_{2}^{(2)}(\ell )+\bar{g}_{1}^{(1)}(\ell )-\kappa (\bar{g}_{2}^{(1)}(\ell )-%
\bar{g}_{1}^{(2)}(\ell )),}  \nonumber \\
{d \over d\ell}%
{ \ln \bar{\chi}}_{{ SS}^{\prime }}^{\kappa} &=&%
{ -\bar{g}_{4}^{(2)}(\ell )-\bar{g}_{4}^{(1)}(\ell ),}  \nonumber \\
{d \over d\ell}{ \ln \bar{\chi}}_{{ TS}^{\prime }}^{\kappa} &=&%
{ -\bar{g}_{4}^{(2)}(\ell )+\bar{g}_{4}^{(1)}(\ell ),}
\end{eqnarray}

with $\bar{g}_{i}^{(j)}=g_{i}^{(j)}/\pi v_{F}$. CDW$^{'}$ and SDW$^{'}$ refer to the anomalous $q\approx 0$ interband situations
while SS$^{'}$ and TS$^{'}$ refer to the $q\approx 2k_{F}$ ones.

We first report on the calculations without the Coulomb interaction, that is
solely with the phonon-mediated effective electron-electron interactions (at
temperatures below the Debye temperature for non-adiabatic interactions\cite
{Caron}). This was done for arbitrary amplitudes but with the sign
constraints given in Eq. (\ref{phonon}). We find no sign of a dominant
superconducting response in a single-walled armchair carbon nanotube. Charge
correlations ($\bar{\chi}_{CDW}^{-,M}$ and $\bar{\chi}_{CDW^{\prime }}^{-,-}$%
) are in all cases the most divergent and open a pseudo-gap which subdues
the superconducting fluctuations. The introduction of Coulomb interactions,
through $u$ and $b$, reinforces this tendency even more.
\begin{figure}
\psfrag{chibar}{$\overline{\chi}$}
\psfrag{SS1}{$\overline{\chi}_{SS}^{-,M}$}
\psfrag{SS2}{$\overline{\chi}_{SS}^{+,M}$}
\psfrag{CDW1}{$\overline{\chi}_{CDW}^{\kappa,M}$}
\psfrag{CDW2}{$\overline{\chi}_{CDW'}^{+,M}$}
\psfrag{Temperature (K)}{Temperature ($\rm K$)}
\includegraphics[scale=1]{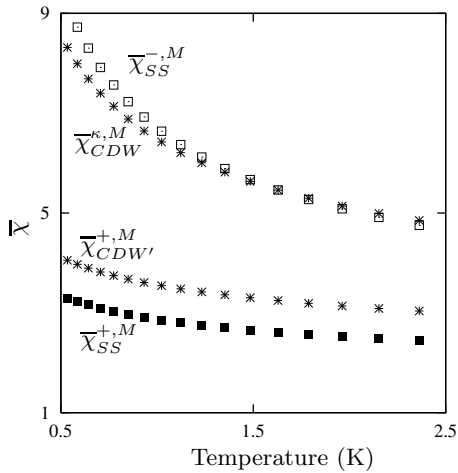}
\caption{\label{armchair} Typical flow diagram for the response functions of a doped armchair
nanotube.}
\end{figure}
Kociak {\it et} {\it al.}\cite{Kociak}, however, mention the possibility
that the band occupancy might not be exactly 1/2. If this were the case, as
far as we can estimate from a tight-binding calculation, the sign
constraints mentioned in Eq.( \ref{phonon}) still hold. However, the Fermi
level would shift by $\Delta E=d v_{F}k_{F}/2$ ($d$ is the doping level)
and the two bands would have different Fermi momenta. As a consequence, for
energy scales below $\Delta E$ the $\bar{g}_{i}^{(3)}$ would vanish because
of longitudinal momentum conservation. Moreover, the interband backward
scattering $g_{1,ph}^{(2)}$, which was previously zero (see Eq.\ref{phonon}%
), now is $\sim d^2g_1$. We now use a three cutoff procedure $
W>\hbar\omega_{ph}>\Delta E$, which applies up to $8\%$ doping level where $\Delta E\approx\hbar\omega_{ph}$. In the case where only the
phonon mediated interactions are considered we have estimated\cite{sedeki} $%
\bar{g}_{1}=-0.3/n$ and $\bar{g}_{2}=\bar{g}_{4}=-0.1/n.$ Superconducting
correlations $\bar{\chi}_{SS}^{-,M}$ are found to dominate for $\ell $ ($=\ln (W/T)$)greater than a critical $\ell _{s}$ ($=\ln (W/T_s)$). This is shown for a typical run in 
Fig.~\ref{armchair} (only those giving the largest enhancement are shown).  Fig.~\ref{phonon} shows the crossover temperatures $T_{s}$ below which this occurs
for various doping levels and different diameter SWCNTs . The turning on of
the Coulomb interaction lowers $T_{s}$ and eventually destroys the
superconducting fluctuations dominance for $u\gtrsim \left| g_{1}\right| $.
\begin{figure}
\psfrag{Temperature (K)}{Temperature ($\rm K$)}
\psfrag{Doping level}{Doping level}
\psfrag{u = 0}{u=0}
\psfrag{b = 0}{b=0}
\includegraphics[scale=1]{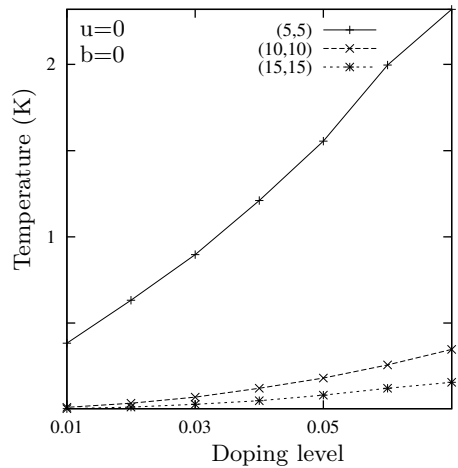}
\caption{\label{phonon} Temperature at which the superconducting fluctuations dominate as a
function of the doping level for three $(n,n)$ armchair tubes. Only
phonon-mediated interactions are considered.}
\end{figure}
At this point, we can conclude that superconducting fluctuations in single
armchair nanotubes originating from a phonon mechanism are possible at an
appreciable doping level provided the Coulomb interaction is very small.
This last condition seems restrictive in view of our own evaluation $u>>\left|g_{1}\right|$ 
and the evidence of strong electronic correlations
in SWCNT \cite{corr}. Moreover, the finite length of the nanotubes results
in a discretization of the energy levels\cite{discrete}. Consequently, the
energy scale in the renormalization group procedure cannot be smaller than
this spacing. We estimate this would occur at best at $1\rm K$ which is still
larger than the temperatures at which the work of Kociak\cite{Kociak} shows
any indication of superconductivity. It thus seems likely that the
superconductivity seen in ropes takes place by a coupling effect between the
nanotubes. At the dimensionality crossover temperature $T_{x}\ \sim\
t_{\bot }/\pi $, where $t_{\bot }$ is the net inter-tube hopping amplitude,
inter-tube hopping becomes coherent \cite{Claude}. Any fluctuating
superconducting pair that might exist will thereafter be able to coherently
tunnel between tubes. This surely occurs much before the SWCNT discrete
spectrum is felt. The single tube superconducting fluctuations need not be
dominant for this to take place. The existence of frustration in ropes of
close packed SWCNT would prevent the further development of bond-order wave
deformations (associated with a Kekule CDW modulation\cite{sedeki}) or
magnetic order and allow superconductivity to develop. Moreover, the
interchain particle-hole interactions that might exist at $T_{x}$ could also
play a role in enhancing superconductivity through a mechanism similar to
the one proposed for organic materials\cite{Claude,Claude1}. All of this is
consistent with the small temperatures observed for superconductivity in
ropes of armchair nanotubes.

The recent discovery of superconducting fluctuations at 15 $\mathop{\rm K}$ 
in a single $(5,0)$ zigzag SWCNT\cite{Tang} thus seems puzzling in view of
the above analysis of armchair nanotubes. In this specific instance, the
very strong $\sigma -\pi $ hybridization due to the very small diameter of
the SWCNT changes the band structure in a dramatic way and makes the tubes
metallic instead of being the expected insulating state\cite{Dressel}. The
band structure might share similarities with the one in a conducting $(6,0)$
zigzag SWNT\cite{Blase}. \ The band structure in the vicinity of the Fermi
level would then show the crossing between a non-degenerate band $(1)$ and
doubly-degenerate bands $(2,3)$ of opposite curvature and having a different
transverse angular momentum. Because of the twofold degeneracy, the Fermi
level no longer lies at the crossing point but is offset so that the bands\ $%
(1)$ and $(2,3)$ have different Fermi momenta. Moreover, the tight symmetry
relationship of the armchair bands exists no longer between $(1)$ and $(2,3)$%
. There are also many different phonons contributing to effective
phonon-mediated electron-electron interactions. Finally, all $\bar{g}%
_{i}^{(3)}$ will vanish because of longitudinal momentum conservation. This,
we believe, is quite sufficient to allow for important superconducting
fluctuations in the zigzag tubes. Purely as an illustration and only for
crude order of magnitude estimates, we used the model developed above with
only two cutoffs $W=\Delta E\sim0.3 \mathop{\rm eV}$\cite{Blase}, $\hbar \omega _{ph}\sim0.166\mathop{\rm eV}$, 
and $n=3$ to account for the smaller tube diameter such that $\bar{g}%
_{1}=-0.1$, $\bar{g}_{2}=-0.03$. We find a dominance of the superconducting
response below room temperature with just the phonons. This is much larger
than for the armchair tubes and might explain the origin of the
superconducting fluctuations observed by Tang {\it et} {\it al.}\cite{Tang}%
. But again adding a Coulomb interaction quickly reduces this temperature.
Superconductivity disappears again for $u\gtrsim \left| g_{1}\right| .$ The
results of Tang would thus indicate an unexpectedly small Coulomb
interaction in zig-zag nanotubes.
A more detailed analysis of the three-band RG will be given elsewhere.
\begin{acknowledgments}
The authors thank D. Senechal for useful comments and discussions on several aspects of this work. We also thank the National Sciences and Engineering Research Council of Canada (NSERC), le Fonds pour la Formation de chercheurs et 
l'Aide \`{a} la Recherche du gouvernement du Quebec (FCAR) and the ``Superconductivity program'' of the Institut Canadien de Recherches Avanc\'{e}es (CIAR), for financial support.
\end{acknowledgments}
\newpage 
\bibliography{sedeki}
\end{document}